# Real interest rates, exchange rates and growth in three emerging economies: an exploratory analysis of Brazil, India and Nigeria

Hugo Spring-Ragain
Centre d'Etudes Diplomatiques et Stratégiques (CEDS)
hugo.springragain@edu.ceds.fr / raguhugo37@gmail.com

**Abstract:**

This paper studies the dynamic relationships between real growth, the real lending interest rate and the exchange rate in Brazil, India and Nigeria over 2000-2022. The theoretical framework places these relationships within the literature on the procyclicality of macroeconomic policies, "fear of floating" and the global financial cycle, without treating the observed variables as monetary policy instruments directly controlled by central banks.

Activity is measured by real GDP in constant local currency (World Bank, NY.GDP.MKTP.KN). Real growth is defined as 100 × Δlog(real GDP) and the change in the exchange rate as 100 × Δlog(exchange rate), with the exchange rate quoted in units of local currency per dollar. The interest rate variable is the World Bank real interest rate (FR.INR.RINR), that is, a lending rate adjusted for inflation as measured by the GDP deflator; it is not a policy rate. Parsimonious bivariate VARs are estimated after examining stationarity, with lag selection by BIC and orthogonalised impulse responses.

In the baseline specification, Brazil shows two pointwise exclusions of zero following a real interest rate innovation: a contraction at one year and a rebound at three years. Neither feature survives all the checks: the rebound disappears in a VAR(1) and is no longer significant when the sample ends in 2019, while the contraction is no longer distinguishable from zero in that same pre-Covid sample. The estimated responses for India and Nigeria remain imprecise, as do the growth responses to exchange rate innovations. Given the very small sample, the large number of horizons examined, the absence of global financial variables and the non-structural nature of the innovations, the results are interpreted as descriptive and exploratory rather than as causal effects of monetary policy.

## Introduction

Emerging economies are exposed simultaneously to fluctuations in domestic activity and to international financial conditions. The crises of 1997, 2008 and 2020 illustrate how funding stress, capital flows and the exchange rate can amplify national cycles. The central difficulty is therefore not only to stabilise activity, but to do so when external constraints may push the authorities to tighten monetary conditions precisely as the economy slows.

This tension lies at the heart of the literature on procyclicality in emerging countries. Kaminsky, Reinhart and Végh (2004) document frequently procyclical monetary policies in these economies, while Végh and Vuletin (2013) show that some of them have gradually acquired a greater countercyclical capacity. Calvo and Reinhart (2002) emphasise the related "fear of floating", and Rey (2013) stresses the role of the global financial cycle in limiting monetary autonomy.

Within this framework, the empirical question addressed is deliberately narrower than an evaluation of monetary policy: to what extent are changes in the real lending interest rate and in the exchange rate associated with real growth dynamics in Brazil, India and Nigeria? The aim is to describe reduced-form dynamics, not to identify a central bank reaction function or a structural monetary policy shock.

## Theoretical and historical background

### *Business cycles and financial crises*

A business cycle refers to the alternation of phases of expansion and slowdown around a trend. In open emerging economies, these fluctuations can be amplified by capital movements, risk premia and the exchange rate; this interaction motivates the joint study of activity, interest rates and the exchange rate.

The literature thus distinguishes internal stabilisation from the external constraint. When capital withdraws during a slowdown, the need to defend the currency or to contain imported inflation can lead to procyclical tightening (Kaminsky, Reinhart and Végh, 2004; Végh and Vuletin, 2013).

Financial crises are episodes in which these mechanisms become particularly visible. They serve here as historical context; the annual VAR does not attempt to identify each crisis separately.

The 2000-2022 period includes, in particular, the global financial crisis and the COVID-19 pandemic. The latter carries considerable weight in the sample and is the subject of a specific robustness check for the Brazilian result.

### *The role of dominant currencies*

The dollar plays a central role in trade invoicing, reserves and financial contracts. The "dominant currency paradigm" shows that this concentration of invoicing can transmit dollar fluctuations beyond bilateral trade alone (Gopinath and Itskhoki, 2022; Bacchetta and van Wincoop, 2005).

An appreciation of the dollar or a depreciation of the domestic currency can raise the cost of imports and of servicing foreign currency liabilities. More broadly, the global financial cycle co-moves with credit conditions and capital flows, which limits monetary policy autonomy even under a flexible exchange rate (Rey, 2013; Druck, Magud and Mariscal, 2018). These channels are not, however, identified separately in the bivariate VARs estimated here.

*The concept of countercyclical monetary policy*

Monetary policy is countercyclical when it tends to ease monetary conditions in periods of weak activity and to tighten them during expansions. In emerging countries, this logic can be thwarted by exchange rate pressures: the need to raise rates to defend the currency during a crisis corresponds to what Végh and Vuletin (2013) call the "fear of free falling". This paper does not directly test this reaction function, since the available interest rate variable is a real lending rate and not a policy rate.

The evolution of monetary frameworks since the 1990s nevertheless suggests that this constraint is not uniform. Some countries have gained credibility and countercyclical capacity, while others remain more exposed to trade-offs between activity, inflation and exchange rate stability (Kaminsky, Reinhart and Végh, 2004; Végh and Vuletin, 2013).

**Constraints on monetary transmission and national experiences**

*Monetary strategies adopted by emerging economies*

In emerging economies, the authorities combine policy rates, liquidity management, foreign exchange intervention, macroprudential instruments and, at times, capital flow management measures. These instruments are neither equivalent nor perfect substitutes; their transmission depends on financial depth and on the structure of the banking system (Mishra and Montiel, 2013; Mishra, Montiel and Spilimbergo, 2012).

*Interest rate policies.*

Policy rates can influence the cost of funding, demand and expectations, but pass-through to lending rates is often incomplete in shallower financial systems. This distinction is essential here, since FR.INR.RINR measures a real lending rate and not the monetary policy rate (Mishra and Montiel, 2013).

The monetary policy reaction also depends on external constraints. When the currency depreciates sharply, a central bank may hesitate to ease despite the slowdown, notably because of the risk of imported inflation or capital outflows; this is one of the mechanisms associated with "fear of floating" (Calvo and Reinhart, 2002).

*Foreign exchange intervention*

Foreign exchange intervention allows a central bank to buy or sell foreign currency in order to provide liquidity or to moderate movements deemed excessive. In economies where dollar invoicing and foreign currency liabilities are large, a depreciation can both support some exporters and raise the cost of imports or tighten balance sheet constraints. The paper does not measure these interventions directly; they are only one element of the institutional framework within which exchange rate movements are interpreted (Gopinath and Itskhoki, 2022; Calvo and Reinhart, 2002).

*Capital controls and other unconventional measures.*

Capital flow management measures can target inflows, outflows or specific categories of transactions, through taxes, prudential requirements or regulatory restrictions. Their function must be distinguished according to the objective pursued: limiting speculative inflows, reducing destabilising outflows or altering the composition of flows. Their evaluation lies beyond the scope of the bivariate VARs in this paper; they are mentioned only as a possible constraint on, or complement to, monetary policy.

*The Brazilian, Indian and Nigerian cases*

The choice of these three countries is exploratory and is not intended to form a representative sample of emerging economies as a whole. It allows a comparison of three large economies located in different regions and characterised by distinct productive structures, exchange rate regimes and degrees of external exposure. This diversity motivates the comparison but does not, on its own, allow the estimated differences to be attributed to any particular institutional feature.

In Brazil, the response to the global crisis combined monetary easing, foreign exchange intervention and, when capital returned in force, measures targeting inflows. In October 2009, the IOF was reintroduced at 2% on foreign purchases of domestic bonds and equities, then raised in 2010 for certain debt flows. These measures targeted capital inflows and appreciation pressure on the real; they should therefore not be described as restrictions intended to directly prevent capital outflows (IMF, 2013).

In India, the Reserve Bank of India eased its policy substantially after the global crisis intensified: between October 2008 and April 2009, the repo rate was cut by 425 basis points, alongside liquidity measures and a reduction in the cash reserve ratio (Reserve Bank of India, 2009). This response fits a framework in which the RBI combines domestic objectives, liquidity management and foreign exchange market intervention (Torija-Zane, 2012).

In Nigeria, macroeconomic policy is closely tied to oil revenues and to the functioning of the foreign exchange market. After the 2014 oil shock, the exchange rate regime combined discrete adjustments, interventions, restrictions and multiple windows. The IMF noted a large parallel market premium in 2017 and, in 2022, several exchange rate windows still coexisted (IMF, 2017; IMF, 2023). The average official rate used in the analysis must therefore be interpreted with caution: it partly reflects an administered price and does not always capture the marginal cost of foreign currency.

*Challenges and limits of countercyclical policies*

*Structural and institutional constraints.*

Structural constraints can weaken monetary transmission: shallow markets, limited access to credit, bank segmentation and institutional quality reduce the ability of a policy rate change to pass through to the financial conditions actually faced by households and firms (Mishra and Montiel, 2013). These mechanisms are particularly important for interpreting the Nigerian results.

Central bank independence, the credibility of the monetary framework and the depth of the financial system also affect the anchoring of expectations and policy choices (Cukierman, Webb and Neyapti, 1992).

*Reactions of financial markets and international investors.*

Finally, global financial conditions can reduce domestic room for manoeuvre. A rise in US interest rates, an increase in risk aversion or an appreciation of the dollar can tighten external financing and increase pressure on the exchange rate (Rey, 2013; Druck, Magud and Mariscal, 2018). These factors are not included in the bivariate VARs; their effects may therefore be absorbed by the estimated innovations, which is an important limitation for interpretation.

## Exploratory econometric analysis

*Data, variables and transformations*

The series in levels cover 2000-2022, i.e. 23 annual observations per country. First-difference transformations yield 22 observations over 2001-2022. After accounting for lags, the one-lag and two-lag models use 21 and 20 effective observations respectively. This sample size requires parsimonious specifications and a cautious interpretation of the coefficients and impulse responses.

The activity variable is real GDP in constant local currency (World Bank, World Development Indicators, NY.GDP.MKTP.KN). This choice avoids the accounting relationship that would arise if GDP were measured in current dollars while the exchange rate against the dollar also appeared in the model.

The other series also come from the World Development Indicators. The real interest rate is the "Real interest rate (%)" (FR.INR.RINR): it is the lending rate adjusted for inflation as measured by the GDP deflator. It is expressed in percent and is not a policy rate; its first difference is therefore expressed in percentage points. The exchange rate is the annual average official rate, in units of local currency per dollar (PA.NUS.FCRF). An increase corresponds to a depreciation of the domestic currency. For Nigeria, the existence of several exchange rate windows and of a large parallel premium during part of the sample makes this official measure particularly imperfect (IMF, 2017; IMF, 2023).

The transformations used are: $g_t = 100 \times \Delta\log(Y_t)$ for real GDP growth; $\Delta i_t$ for the change in the real interest rate; and $\Delta e_t = 100 \times \Delta\log(E_t)$ for the change in the exchange rate. Thus, $\Delta e_t > 0$ represents a depreciation of the domestic currency against the dollar. Inflation is not included in the bivariate VARs presented and is therefore no longer used as an empirical variable in this version.

*General model specification*

For each country, the main econometric analysis relies on bivariate vector autoregressive models in the tradition of Sims (1980). The aim is to study the dynamic relationships between real GDP growth and, alternatively, the change in the real interest rate or the change in the exchange rate.

The vector of endogenous variables is defined as:

$$z_t = \begin{pmatrix} g_t \\ x_t \end{pmatrix},$$

where $g_t$ is real GDP growth and $x_t$ denotes, alternatively:

$$x_t = \Delta i_t$$

for the growth–real interest rate model, or:

$$x_t = \Delta e_t$$

for the growth–exchange rate model.

The VAR model of order $p$, denoted VAR($p$), is written:

$$z_t = c + A_1 z_{t-1} + A_2 z_{t-2} + \cdots + A_p z_{t-p} + u_t,$$

or, in compact form:

$$z_t = c + \sum_{j=1}^{p} A_j \, z_{t-j} + u_t.$$

The vector of constants is:

$$c = \begin{pmatrix} c_g \\ c_x \end{pmatrix},$$

and, for each lag $j$, the matrix of autoregressive coefficients is:

$$A_j = \begin{pmatrix} a_{11,j} & a_{12,j} \\ a_{21,j} & a_{22,j} \end{pmatrix}.$$

The vector of reduced-form innovations is:

$$u_t = \begin{pmatrix} u_{g,t} \\ u_{x,t} \end{pmatrix},$$

with:

$$\mathbb{E}(u_t) = 0, \qquad \mathbb{E}(u_t u_t{}') = \Sigma_u,$$

and:

$$\mathbb{E}(u_t u_s{}') = 0 \qquad \text{for } t \neq s.$$

The variance-covariance matrix of the innovation process is:

$$\Sigma_u = \begin{pmatrix} \sigma_g^2 & \sigma_{gx} \\ \sigma_{gx} & \sigma_x^2 \end{pmatrix}.$$

The covariance $\sigma_{gx}$ means that the reduced-form innovations of the two equations may be contemporaneously correlated. This correlation motivates the orthogonalisation used for the impulse response functions.

*Equation-by-equation form*

The bivariate system is equivalent to:

$$g_t = c_g + \sum_{j=1}^{p} a_{11,j}\, g_{t-j} + \sum_{j=1}^{p} a_{12,j}\, x_{t-j} + u_{g,t},$$

$$x_t = c_x + \sum_{j=1}^{p} a_{21,j}\, g_{t-j} + \sum_{j=1}^{p} a_{22,j}\, x_{t-j} + u_{x,t}.$$

In the case of a VAR(1):

$$\begin{pmatrix} g_t \\ x_t \end{pmatrix} = \begin{pmatrix} c_g \\ c_x \end{pmatrix} + \begin{pmatrix} a_{11,1} & a_{12,1} \\ a_{21,1} & a_{22,1} \end{pmatrix} \begin{pmatrix} g_{t-1} \\ x_{t-1} \end{pmatrix} + \begin{pmatrix} u_{g,t} \\ u_{x,t} \end{pmatrix}.$$

In the case of a VAR(2):

$$z_t = c + A_1 z_{t-1} + A_2 z_{t-2} + u_t.$$

*Variable transformations*

Real GDP growth is defined as:

$$g_t = 100[\ln(Y_t) - \ln(Y_{t-1})] = 100\,\Delta\ln(Y_t),$$

where $Y_t$ denotes real GDP in constant local currency.

The change in the real interest rate is:

$$\Delta i_t = i_t - i_{t-1}.$$

Since the rate $i_t$ is expressed in percent, $\Delta i_t$ is expressed in percentage points.

The change in the exchange rate is:

$$\Delta e_t = 100[\ln(E_t) - \ln(E_{t-1})] = 100\,\Delta\ln(E_t),$$

where $E_t$ is the number of units of local currency per US dollar. Thus:

$$\Delta e_t > 0 \Rightarrow \text{depreciation of the domestic currency},$$

$$\Delta e_t < 0 \Rightarrow \text{appreciation of the domestic currency}.$$

Inflation does not enter the core bivariate VARs of the final version and is therefore not included in the main formalisation.

*VAR estimation and two distinct estimators of the residual covariance*

Stacking the observations used in estimation, the system can be written:

$$\mathbf{Z} = \mathbf{X}B + U,$$

where:

- $\mathbf{Z}$ has dimension $T \times K$;
- $\mathbf{X}$ has dimension $T \times (Kp + 1)$ when a constant is included;
- $B$ has dimension $(Kp + 1) \times K$;
- $U$ has dimension $T \times K$;
- here $K = 2$.

The OLS estimator is:

$$\hat{B} = (\mathbf{X}'\mathbf{X})^{-1}\mathbf{X}'\mathbf{Z}.$$

Since both equations of the VAR contain exactly the same regressors, equation-by-equation OLS is equivalent to SUR estimation and is efficient under the usual VAR assumptions.

The estimated residuals are:

$$\hat{U} = \mathbf{Z} - \mathbf{X}\hat{B},$$

or, observation by observation:

$$\hat{u}_t = z_t - \hat{c} - \sum_{j=1}^{p} \hat{A}_j\, z_{t-j}.$$

Two estimators of the residual covariance must be distinguished, because *statsmodels* does not use them for the same objects.

*Maximum likelihood estimator used in the BIC*

The matrix used in the BIC determinant is:

$$\hat{\Sigma}_u^{ML} = \frac{1}{T}\sum_{t=1}^{T} \hat{u}_t\, \hat{u}_t'.$$

*Corrected estimator used for the IRFs and the bootstrap simulation*

With a constant and $K$ endogenous variables, each equation contains $Kp + 1$ parameters. The degrees-of-freedom-corrected estimator is:

$$\hat{\Sigma}_u = \frac{1}{T - Kp - 1}\sum_{t=1}^{T} \hat{u}_t\, \hat{u}_t'.$$

It is this matrix $\hat{\Sigma}_u$ that *statsmodels* uses for the Cholesky decomposition of the IRFs in the replication pipeline, and that the final script uses to simulate the innovations of the parametric bootstrap.

In the Brazilian VAR(2), $T = 20$, $K = 2$ and $p = 2$, so that:

$$T - Kp - 1 = 20 - 4 - 1 = 15.$$

The distinction between $\hat{\Sigma}_u^{ML}$ and $\hat{\Sigma}_u$ is therefore quantitatively important in this very small sample.

*Lag selection by BIC*

The Bayesian information criterion used by *statsmodels* is:

$$BIC(p) = \ln\left|\hat{\Sigma}_{u,p}^{ML}\right| + \frac{\ln(T)}{T} k_p.$$

With $K$ variables and a constant, the total number of penalised free parameters is:

$$k_p = K^2 p + K.$$

In a bivariate VAR ($K = 2$):

$$k_p = 4p + 2.$$

Thus, for $p = 2$:

$$k_2 = 4 \times 2 + 2 = 10.$$

The selected lag is:

$$\hat{p} = \underset{p \in \{1,2\}}{\operatorname{argmin}}\ BIC(p).$$

In the replication pipeline, the BICs of the VAR(1) and VAR(2) are computed over a common window so that the competing models are compared on the same number of dependent observations. The selected model is then re-estimated on the maximum number of observations compatible with $\hat{p}$.

*Stability condition*

The VAR($p$) can be written in companion form:

$$\mathcal{Z}_t = \mathcal{C} + F\mathcal{Z}_{t-1} + \mathcal{U}_t,$$

with:

$$\mathcal{Z}_t = \begin{pmatrix} z_t \\ z_{t-1} \\ \vdots \\ z_{t-p+1} \end{pmatrix},$$

and:

$$F = \begin{pmatrix} A_1 & A_2 & \cdots & A_{p-1} & A_p \\ I & 0 & \cdots & 0 & 0 \\ 0 & I & \cdots & 0 & 0 \\ \vdots & \vdots & \ddots & \vdots & \vdots \\ 0 & 0 & \cdots & I & 0 \end{pmatrix}.$$

The VAR is stable if all eigenvalues $\lambda_r$ of $F$ satisfy:

$$|\lambda_r| < 1 \qquad \forall r.$$

An equivalent formulation relies on the polynomial:

$$A(z) = I - A_1 z - \cdots - A_p z^p.$$

Stability requires that the roots of:

$$\det\left(I - A_1 z - \cdots - A_p z^p\right) = 0$$

lie outside the unit circle, under the root convention of the autoregressive polynomial.

*Unconditional mean of the VAR*

Under stability, $I - A_1 - \cdots - A_p$ is invertible. The unconditional mean $\mu$ of the process is:

$$\mu = \left(I - A_1 - \cdots - A_p\right)^{-1} c.$$

This definition allows the moving average representation to be written around the unconditional mean.

*Recursive identification by Cholesky decomposition*

We introduce a vector of orthogonal structural shocks:

$$\varepsilon_t = \begin{pmatrix} \varepsilon_{g,t} \\ \varepsilon_{x,t} \end{pmatrix},$$

with:

$$\mathbb{E}(\varepsilon_t) = 0, \qquad \mathbb{E}(\varepsilon_t \varepsilon_t{}') = I.$$

The reduced-form innovations are related to the orthogonalised shocks by:

$$u_t = \Lambda \varepsilon_t,$$

where $\Lambda$ is the unique lower-triangular Cholesky factor with strictly positive diagonal satisfying:

$$\hat{\Sigma}_u = \Lambda \Lambda'.$$

In the main ordering:

$$z_t = \begin{pmatrix} g_t \\ x_t \end{pmatrix},$$

the factor is:

$$\Lambda = \begin{pmatrix} \lambda_{11} & 0 \\ \lambda_{21} & \lambda_{22} \end{pmatrix}.$$

Thus:

$$u_{g,t} = \lambda_{11} \varepsilon_{g,t},$$

$$u_{x,t} = \lambda_{21} \varepsilon_{g,t} + \lambda_{22} \varepsilon_{x,t}.$$

The recursive restriction implies:

$$\frac{\partial g_t}{\partial \varepsilon_{x,t}} = 0,$$

whereas:

$$\frac{\partial x_t}{\partial \varepsilon_{g,t}} = \lambda_{21}.$$

In the main ordering, an orthogonalised shock to $x_t$ therefore does not affect growth contemporaneously, whereas a growth shock can contemporaneously affect the variable placed second.

*Reverse ordering and permutation matrix*

To check the sensitivity of the results to the Cholesky ordering, we define the permutation matrix:

$$\Pi = \begin{pmatrix} 0 & 1 \\ 1 & 0 \end{pmatrix}.$$

The reordered vector is:

$$z_t^{(\Pi)} = \Pi z_t.$$

The autoregressive matrices and the covariance become:

$$A_j^{(\Pi)} = \Pi A_j \Pi',$$

$$\Sigma_u^{(\Pi)} = \Pi \hat{\Sigma}_u \Pi'.$$

We then compute:

$$\Lambda_\Pi = \text{chol}\left(\Sigma_u^{(\Pi)}\right),$$

with $\Lambda_\Pi$ lower triangular with positive diagonal. The responses are computed in the permuted system and then returned to the original ordering by multiplication by $\Pi'$.

*Moving average representation*

Under stability:

$$z_t = \mu + \sum_{h=0}^{\infty} \Psi_h \, u_{t-h}.$$

With:

$$\Psi_0 = I,$$

and the recursion:

$$\Psi_h = \sum_{j=1}^{\min(p,h)} A_j \, \Psi_{h-j},$$

with $\Psi_h = 0$ for $h < 0$.

For a VAR(1):

$$\Psi_h = A_1^h.$$

Using $u_t = \Lambda \varepsilon_t$:

$$z_t = \mu + \sum_{h=0}^{\infty} \Psi_h \, \Lambda \varepsilon_{t-h}.$$

Defining:

$$\Theta_h = \Psi_h \Lambda,$$

we obtain:

$$z_t = \mu + \sum_{h=0}^{\infty} \Theta_h \, \varepsilon_{t-h}.$$

*Impulse response functions*

We define the selection vectors:

$$s_1 = \begin{pmatrix} 1 \\ 0 \end{pmatrix}, \qquad s_2 = \begin{pmatrix} 0 \\ 1 \end{pmatrix}.$$

The response of growth to an orthogonalised shock to the second variable at horizon $h$ is:

$$IRF_{g \leftarrow x}(h) = s_1{}' \Theta_h s_2 = s_1{}' \Psi_h \Lambda s_2.$$

It corresponds to:

$$IRF_{g \leftarrow x}(h) = \frac{\partial g_{t+h}}{\partial \varepsilon_{x,t}}.$$

In the main ordering:

$$IRF_{g \leftarrow x}(0) = s_1{}' \Lambda s_2 = 0.$$

The reverse response is:

$$IRF_{x \leftarrow g}(h) = s_2{}' \Psi_h \Lambda s_1.$$

This reverse response describes the dynamic association between a growth innovation and the interest rate or exchange rate variable; it does not constitute a structural central bank reaction function.

*Size of the orthogonalised shock*

The normalisation $\mathbb{E}(\varepsilon_t \varepsilon_t{}') = I$ means that the structural shock has unit variance. In the main ordering, a unit shock to $\varepsilon_{x,t}$ produces a contemporaneous change in $x_t$ equal to:

$$\lambda_{22}.$$

$\lambda_{22}$ is therefore the size, in the units of $x_t$, of the orthogonalised shock shown in the IRFs.

For the Brazil growth–real interest rate model in VAR(2):

$$\lambda_{22} \simeq 5.77$$

percentage points of change in the real lending interest rate. A growth response of $-1.67$ point at the one-year horizon thus corresponds, as a purely descriptive rescaling, to approximately:

$$\frac{-1.67}{5.77} \simeq -0.29$$

point of growth per point increase in $\Delta i_t$. This conversion does not turn the innovation into a causal monetary policy shock.

For reference, the values of $\lambda_{22}$ obtained in the six core specifications are approximately:

| Country | Model | Lag | $\lambda_{22}$ |
|---|---|---|---|
| Brazil | growth–$\Delta i$ | 2 | 5.77 |
| Brazil | growth–$\Delta e$ | 1 | 10.17 |
| India | growth–$\Delta i$ | 1 | 2.40 |
| India | growth–$\Delta e$ | 1 | 5.65 |
| Nigeria | growth–$\Delta i$ | 1 | 6.11 |
| Nigeria | growth–$\Delta e$ | 1 | 8.62 |

For $\Delta i$, the unit is the percentage point. For $\Delta e$, the unit corresponds approximately to a percentage change in the exchange rate, since $\Delta e_t = 100\Delta \ln(E_t)$.

*Parametric bootstrap of the IRFs*

From the estimated VAR:

$$z_t = \hat{c} + \sum_{j=1}^{p} \hat{A}_j \, z_{t-j} + \hat{u}_t,$$

the replication pipeline uses the corrected covariance:

$$\hat{\Sigma}_u = \frac{1}{T - Kp - 1} \sum_{t=1}^{T} \hat{u}_t \, \hat{u}_t'.$$

For each replication $r = 1, \dots, R$, with:

$$R = 1000,$$

artificial innovations are drawn according to:

$$u_t^{*(r)} \sim \mathcal{N}(0, \hat{\Sigma}_u).$$

An artificial path is generated from the estimated coefficients:

$$z_t^{*(r)} = \hat{c} + \sum_{j=1}^{p} \hat{A}_j \, z_{t-j}^{*(r)} + u_t^{*(r)}.$$

In the final implementation, the simulation uses a *burn-in* period of 100 observations:

$$H_{burn} = 100.$$

The first 100 simulated observations are discarded before re-estimation. This procedure reduces the dependence of the retained bootstrap sample on the simulator's internal initial conditions.

The number of lags is held fixed at the value $\hat{p}$ selected on the original sample: it is not re-selected by BIC in each replication. The bootstrap therefore quantifies uncertainty conditional on the selected model and **does not incorporate the uncertainty arising from lag selection**.

The VAR is re-estimated on each path:

$$z_t^{*(r)} = \hat{c}^{*(r)} + \sum_{j=1}^{p} \hat{A}_j^{*(r)} z_{t-j}^{*(r)} + \hat{u}_t^{*(r)},$$

and the corresponding IRF is computed:

$$IRF^{*(r)}(h).$$

In the main procedure, unstable re-estimates are discarded and simulations continue until $R = 1000$ accepted replications are obtained. The bands are therefore conditional on this stability filter.

For each horizon $h$, the 95% pointwise interval is:

$$CI_{0.95}(h) = [Q_{0.025}\{IRF^*(h)\}, Q_{0.975}\{IRF^*(h)\}].$$

These intervals are pointwise, not simultaneous across all horizons.

*Augmented Dickey–Fuller test*

For a generic series $y_t$, the ADF auxiliary regression with a constant is:

$$\Delta y_t = \alpha + \gamma y_{t-1} + \sum_{j=1}^{q} \delta_j \Delta y_{t-j} + \eta_t.$$

The null hypothesis is:

$$H_0: \gamma = 0,$$

i.e. the presence of a unit root. The alternative is:

$$H_1: \gamma < 0.$$

The ADF statistic does not follow a standard Student distribution under $H_0$. P-values and critical values are computed from the MacKinnon approximations used by *statsmodels*.

Given the small sample size, the maximum number of lags is limited to:

$$q_{max} = 2,$$

and the effective lag is then chosen by AIC among the admissible specifications.

*KPSS test*

With a constant, the residuals used are:

$$\hat{r}_t = y_t - \bar{y}.$$

The partial sums are:

$$S_t = \sum_{j=1}^{t} \hat{r}_j.$$

The KPSS statistic is:

$$KPSS = \frac{T^{-2} \sum_{t=1}^{T} S_t^2}{\widehat{\omega}^2}.$$

$\widehat{\omega}^2$ is a Newey–West estimator of the long-run variance of the residuals. In the replication pipeline, *statsmodels* is called with *nlags="auto"*: the truncation parameter is chosen by the automatic procedure based on Hobijn et al., as implemented in *statsmodels*.

The null hypothesis is stationarity around a constant:

$$H_0\colon\ y_t \text{ is stationary around a constant,}$$

against:

$$H_1\colon\ y_t \text{ is not stationary around a constant.}$$

A joint reading of ADF and KPSS is preferred given the small sample size.

*Autocorrelation diagnostics and multivariate whiteness test*

The white noise assumptions concern the theoretical innovation process $u_t$:

$$\mathbb{E}(u_t) = 0,$$

and:

$$\Gamma_u(j) = \mathbb{E}(u_t u_{t-j}') = 0 \qquad \text{for } j \geq 1.$$

The empirical tests are built from the autocovariances of the estimated residuals:

$$\hat{C}_j = \frac{1}{T} \sum_{t=j+1}^{T} \hat{u}_t\, \hat{u}_{t-j}'.$$

The adjusted multivariate portmanteau statistic used to test whiteness up to horizon $h$ can be written:

$$Q_h^* = T^2 \sum_{j=1}^{h} \frac{1}{T-j} \operatorname{tr}\left(\hat{C}_j' \hat{C}_0^{-1} \hat{C}_j \hat{C}_0^{-1}\right).$$

Under the null hypothesis of no autocorrelation, asymptotically:

$$Q_h^* \overset{a}{\sim} \chi^2_{K^2(h-p)}.$$

In a bivariate VAR with $K = 2$, $p = 2$ and a test up to $h = 3$:

$$K^2(h-p) = 4(3-2) = 4$$

degrees of freedom only, which warrants a cautious interpretation in a very short sample.

Equation-by-equation Ljung–Box tests provide a univariate complement; they do not replace the multivariate diagnostic of the system.

*Scope of the identification*

The recursive decomposition constructs orthogonal innovations:

$$\varepsilon_t = \Lambda^{-1} u_t,$$

but this orthogonalisation is not sufficient to establish that the interest rate innovation is an exogenous monetary policy shock.

In this study, the interest rate variable is the World Bank real lending interest rate and not the policy rate set by the central bank. The IRFs are therefore interpreted as dynamic associations conditional on the chosen specification, not as identified causal effects of a monetary policy decision.

Likewise, the exchange rate is the annual official rate against the dollar. In economies where the foreign exchange market is administered or segmented, this series may not fully reflect market conditions. This caveat is particularly important in the Nigerian case.

Finally, the very small sample size means that statistical uncertainty, lag selection uncertainty and sensitivity to extreme episodes must remain central to the interpretation.

*Stationarity, lags and stability*

In line with the preceding formalisation, stationarity is examined jointly using the Augmented Dickey-Fuller (ADF) and KPSS tests, with a constant. Given the 22 observations available after transformation, the ADF test is deliberately restricted to a maximum of two lags; among these parsimonious specifications, the number of lags is chosen by AIC. The KPSS test uses the automatic lag selection of the chosen implementation. Table A1 reports the statistic, the p-value, the number of lags actually retained and the number of observations used. For Nigerian growth, the ADF test does not reject the unit root ($p = 0.148$) and KPSS rejects stationarity around a constant ($p = 0.018$). As a sensitivity check, a Zivot-Andrews test allowing for an endogenous break rejects the unit root ($p = 0.005$) and places the break around 2015; given the very small sample, this result is interpreted as an indication of a possible structural break, not as a definitive validation of stationarity. The Nigerian results are therefore presented with heightened caution.

Given the small sample size, the number of lags is selected by BIC among the one-lag and two-lag specifications. To ensure a consistent comparison, the information criteria are computed over a common window: the VAR(1) is truncated by one initial observation at the selection stage, so that the VAR(1) and VAR(2) are compared on the same number of dependent observations. The selected model is then re-estimated on the full transformed sample available. The six bivariate models retained are stable: two lags for the Brazil growth-real interest rate model, and one lag for the other five models. Equation-by-equation Ljung-Box tests at lags 1 to 3 are complemented by a multivariate residual whiteness test with a small-sample correction. The multivariate test does not reject the absence of autocorrelation in any of the six models at the 5% level, but the univariate Ljung-Box test flags a warning in the $\Delta i$ equation of the Nigerian model at lag 3 ($p = 0.009$). This divergence warrants additional caution for Nigeria.

The impulse responses are orthogonalised by Cholesky decomposition. In the main specification, growth is ordered before the shock variable: the real interest rate or exchange rate innovation therefore has no contemporaneous effect on growth, but can affect it at subsequent horizons; the reverse ordering is estimated as a sensitivity check. Each impulse corresponds to a one-standard-deviation orthogonalised innovation. The 95% pointwise intervals come from a parametric bootstrap with 1,000 accepted replications, with filtering of unstable re-estimated VARs and export of the rejection rate. With a number of effective observations close to 20, this procedure remains fragile: Kilian (1998) shows that bias-corrected bootstrap intervals can improve coverage in small samples. A bootstrap-after-bootstrap procedure is not implemented here; the bands should therefore be read as exploratory measures of uncertainty.

## Results: Brazil

*Real growth and the real interest rate*

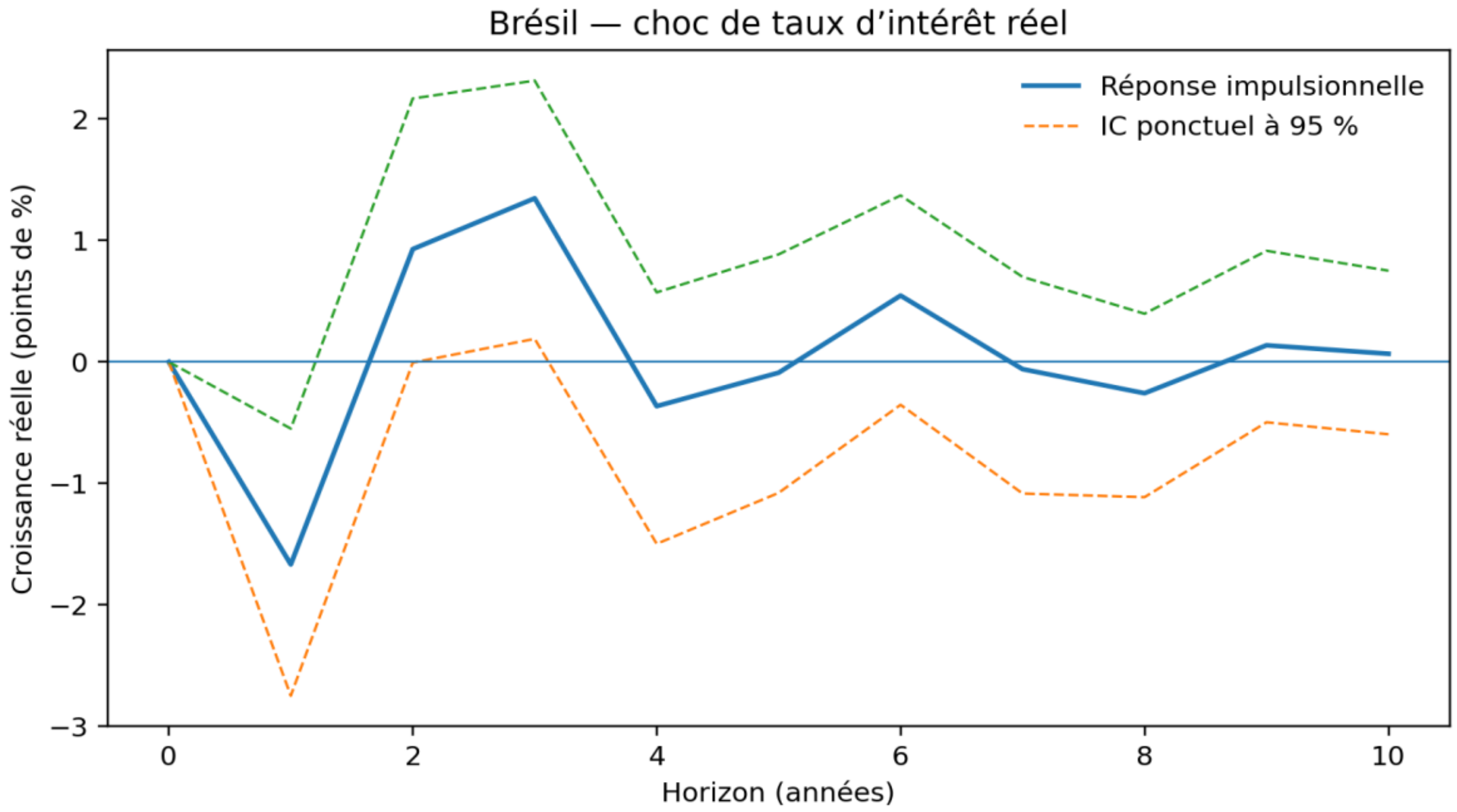

*Figure 1 - Brazil: response of real growth to a real interest rate change shock (main ordering: growth then interest rate; one-standard-deviation innovation; 95% pointwise CI, 1,000 replications)*

In the baseline VAR(2) specification, the response of real growth to a positive real interest rate innovation is negative at one year (-1.67 points; 95% pointwise CI [-2.75, -0.55]) and then positive at three years (+1.35; [0.19, 2.32]). These exclusions of zero are pointwise, not simultaneous, and must be confronted with the robustness checks below before any economic interpretation.

Since the interest rate variable is a real lending rate and not a policy rate, this dynamic cannot be equated with the causal effect of a tightening decided by the central bank. It may also reflect changes in bank spreads, in inflation and in common macro-financial conditions.

**Robustness of the Brazilian result**

Two checks make it possible to distinguish two sources of fragility. First, when a VAR(1) is imposed on the full sample, the one-year contraction remains (-1.54; 95% pointwise CI [-2.80, -0.24]) but the three-year rebound disappears (+0.02; [-0.54, 0.62]): the rebound therefore depends on the second lag. Second, when the VAR(2) is kept but the sample ends in 2019, the one-year response falls to -0.61 and its interval [-1.43, 0.15] contains zero, while the three-year point estimate remains positive (+1.36) but its interval [-0.07, 2.32] also includes zero. The short-term contraction therefore depends heavily on the inclusion of the Covid period, while the three-year rebound depends on the second lag and also loses significance without the Covid period. These checks were recomputed with the same filtered parametric bootstrap as the main results.

These sensitivities matter because the main estimate rests on only 20 effective observations. They lead us to treat the two pointwise exclusions of zero as suggestive results rather than as the central finding of the paper.

*Table 1 - Sensitivity of the Brazilian responses to lags and to the estimation period*

| Specification | Lags | Effective N | IRF h=1 (95% CI) | IRF h=3 (95% CI) |
|---|---|---|---|---|
| Baseline 2000-2022 | 2 | 20 | -1.67 [-2.75, -0.55] | +1.35 [0.19, 2.32] |
| Robustness VAR(1), 2000-2022 | 1 | 21 | -1.54 [-2.80, -0.24] | +0.02 [-0.54, 0.62] |
| Robustness pre-Covid, ending 2019 | 2 | 17 | -0.61 [-1.43, 0.15] | +1.36 [-0.07, 2.32] |

**Real growth and the exchange rate**

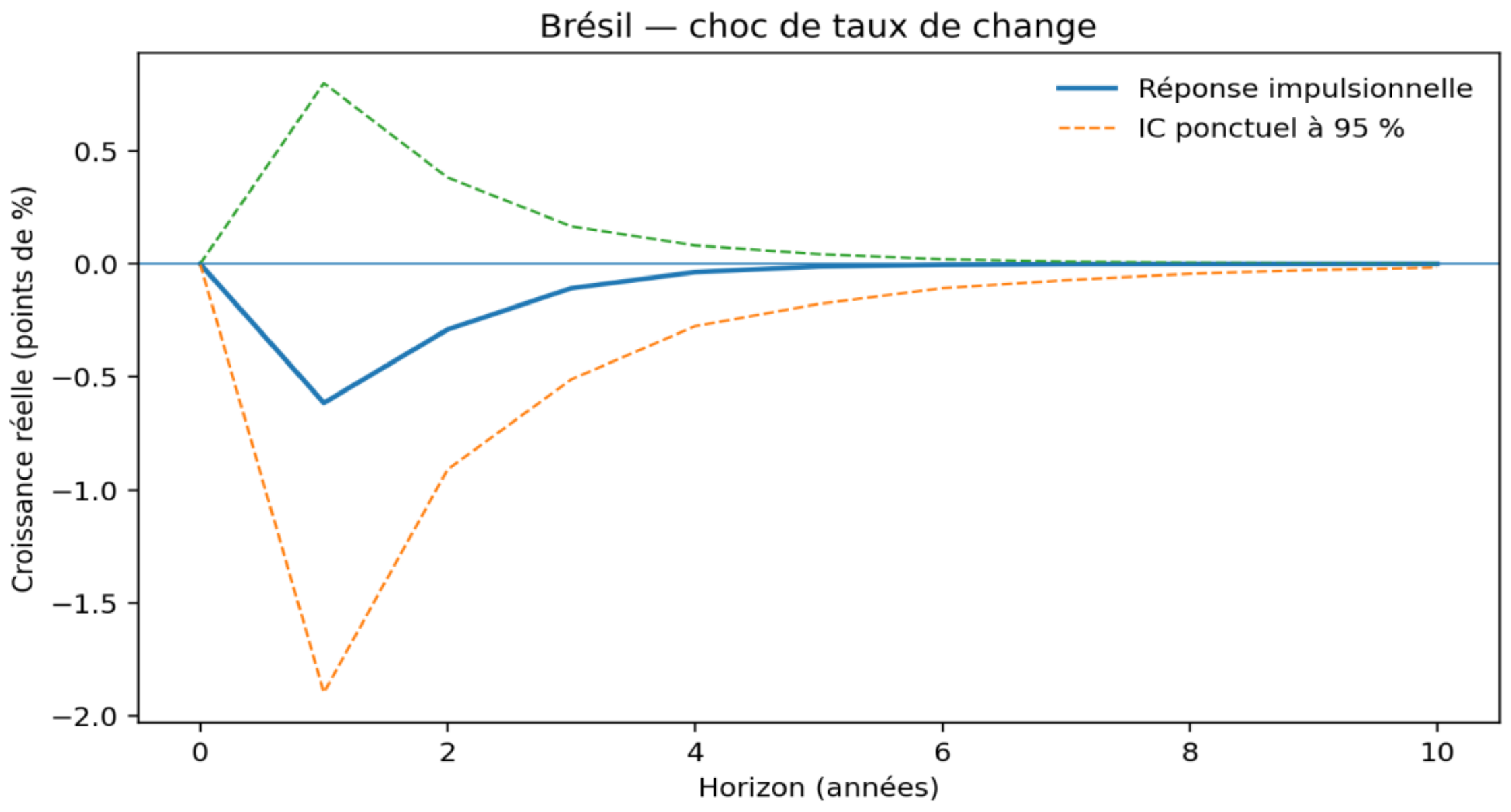

*Figure 2 - Brazil: response of real growth to an exchange rate depreciation shock (main ordering: growth then exchange rate; one-standard-deviation innovation; 95% pointwise CI, 1,000 replications)*

The response of Brazilian growth to a depreciation is negative at the one-year horizon (-0.62 points; 95% pointwise interval [-1.90, 0.80]) and then quickly converges to zero. The estimate does not allow this response to be distinguished from zero at the 5% level. With the reverse Cholesky ordering, a negative contemporaneous response becomes mechanically possible since the exchange rate is ordered first, but the lagged responses remain imprecise; the conclusion regarding future horizons is therefore unchanged.

## Results: India

*Real growth and the real interest rate*

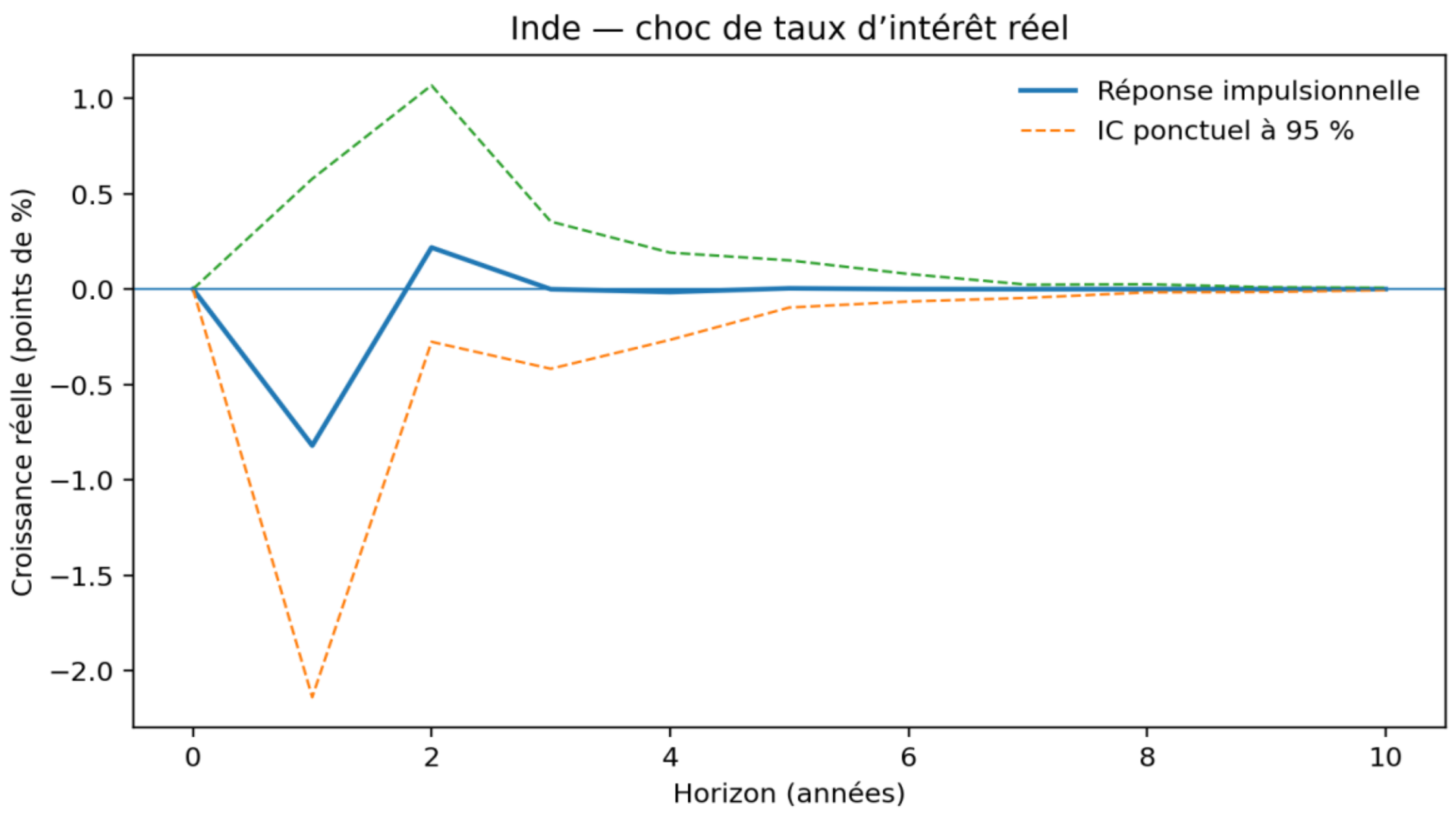

*Figure 3 - India: response of real growth to a real interest rate change shock (main ordering: growth then interest rate; one-standard-deviation innovation; 95% pointwise CI, 1,000 replications)*

In India, the real interest rate innovation is followed by a negative growth response at the one-year horizon (-0.82 points), with a 95% pointwise interval of [-2.14, 0.58]. Subsequent responses are small and quickly converge to zero. The estimate does not allow the response to be distinguished from zero at the chosen level, and the reverse Cholesky ordering does not alter this conclusion.

*Real growth and the exchange rate*

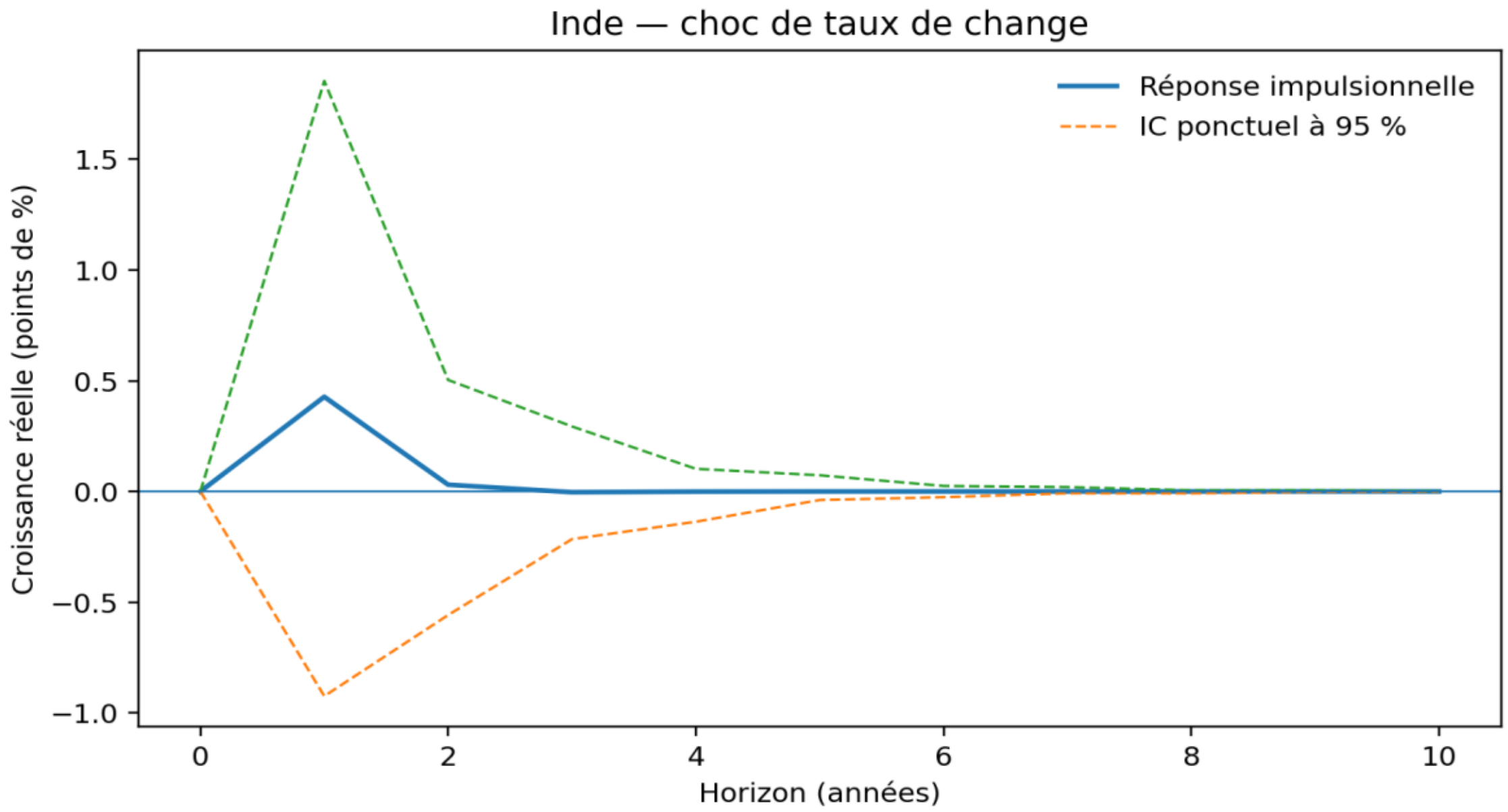

*Figure 4 - India: response of real growth to an exchange rate depreciation shock (main ordering: growth then exchange rate; one-standard-deviation innovation; 95% pointwise CI, 1,000 replications)*

The response of Indian growth to a depreciation is slightly positive at the one-year horizon (+0.43 points), but it is very imprecise: the 95% pointwise interval is [-0.92, 1.85]. It then becomes practically zero. The estimate therefore does not establish a robust sequence of contraction followed by recovery after an exchange rate innovation; the reverse-ordering check leads to the same conclusion.

**Results: Nigeria**

*Real growth and the real interest rate*

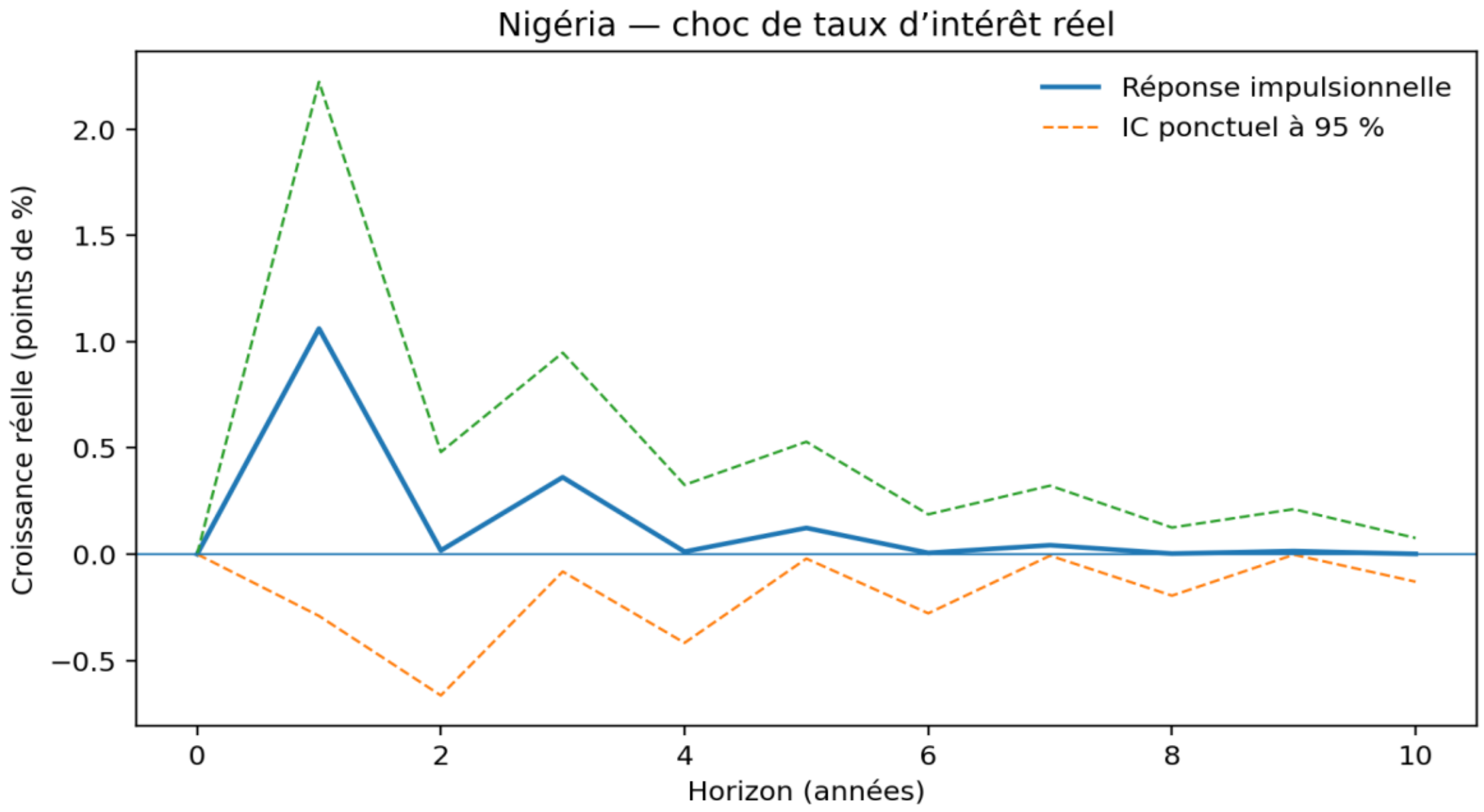

*Figure 5 - Nigeria: response of real growth to a real interest rate change shock (main ordering: growth then interest rate; one-standard-deviation innovation; 95% pointwise CI, 1,000 replications)*

In Nigeria, the response of growth to a positive real interest rate innovation is positive at the one-year horizon (+1.06 points), but the 95% pointwise interval [-0.29, 2.22] includes zero.

Later responses are also imprecise. The interpretation is particularly fragile because of the stationarity diagnostics for growth and the Ljung-Box warning in the Δi equation. The reverse Cholesky ordering does not reveal any lagged response statistically distinguishable from zero.

*Real growth and the exchange rate*

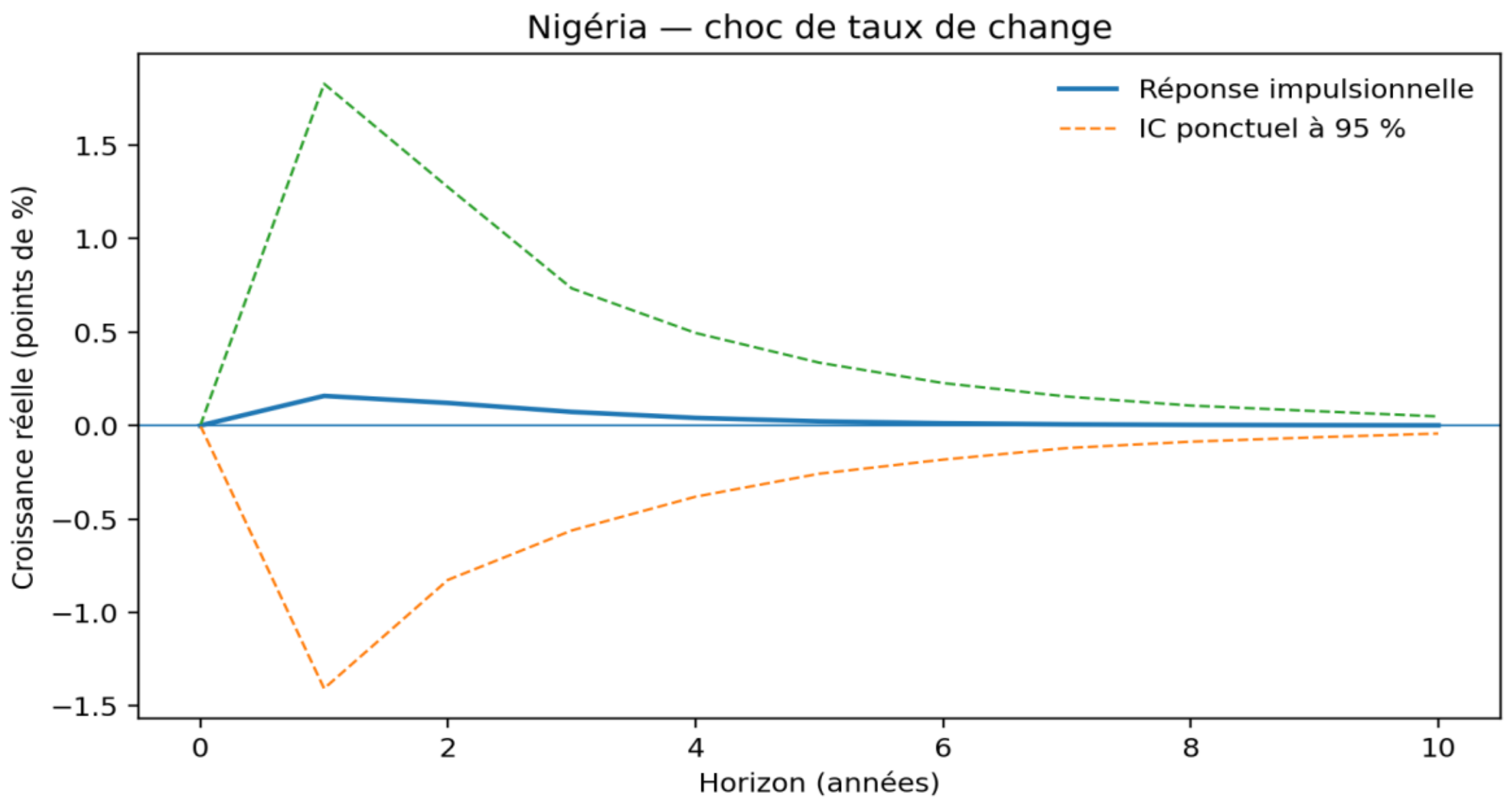

*Figure 6 - Nigeria: response of real growth to an exchange rate depreciation shock (main ordering: growth then exchange rate; one-standard-deviation innovation; 95% pointwise CI, 1,000 replications)*

The IRF associated with the exchange rate is small and positive at the first horizon (+0.16 points), with a 95% pointwise interval of [-1.41, 1.83], and then declines gradually. The bands contain zero at all lagged horizons. The reverse ordering changes the sign of some point estimates, notably because it allows a contemporaneous effect of the exchange rate on growth, but the intervals remain wide. Given this sensitivity and the stationarity diagnostics, this Nigerian result is treated as particularly fragile.

**Discussion and limits of identification**

The point estimates show different profiles across countries, but no general conclusion about transmission survives all the checks. Brazil yields two pointwise exclusions of zero in the baseline VAR(2), but the three-year rebound disappears in the VAR(1) and the one-year contraction is no longer distinguishable from zero in the sample ending in 2019. The exchange rate innovations remain imprecise in all three countries.

Inference is also exposed to a multiplicity problem: six models are examined over roughly ten future horizons each, i.e. on the order of sixty 95% pointwise intervals. Without a correction for multiple comparisons or simultaneous bands, a few exclusions of zero can appear by chance when many horizons are inspected. The two exclusions observed in the baseline Brazilian VAR should therefore not be read as a global validation of the trajectory. Added to this are the very small sample, the problematic stationarity of Nigerian growth, the Cholesky convention and the limitations of the standard parametric bootstrap in small samples (Kilian, 1998).

Using real GDP in constant local currency eliminates the accounting link created by GDP in current dollars, but it does not address the omission of global shocks. The VARs contain no US interest rate, no dollar index and no measure of risk aversion; for Nigeria, they also omit the oil price. The innovations may therefore capture unobserved common shocks. The Nigerian official rate is, moreover, an imperfect indicator of the effective price of foreign currency during periods of market segmentation.

Reverse responses and the link with countercyclicality

The growth-interest rate VARs also contain the response of the real lending interest rate to a growth innovation. At the one-year horizon, the point estimates are +0.51 points in Brazil (95% pointwise CI [-2.31, 2.82]), +0.72 in India ([-0.26, 1.85]) and +0.56 in Nigeria ([-2.45, 3.04]). In all three cases, the intervals include zero by a wide margin: none of these responses is statistically distinguishable from zero at the 5% level. There is therefore no common positive profile that could be interpreted as a countercyclical reaction. In Brazil, the point response in fact becomes negative at two years (-2.51; [-5.71, 0.36]). These values were recomputed with the same filtered parametric bootstrap procedure as the main IRFs. Finally, even a significant response would not constitute a central bank reaction function, since the observed variable is a real lending rate that incorporates bank spreads and the GDP deflator.

The reaction of the interest rate to a depreciation cannot be identified from the two separate bivariate VARs used here. Such a test would require at least a joint system including growth, the interest rate and the exchange rate or, more directly, the estimation of a reaction function with a policy rate. The reverse responses presented above are therefore purely descriptive and, in the sample studied, statistically imprecise.

## Conclusion

The theoretical framework recalls that emerging economies can be constrained by capital flows, the exchange rate and financial depth when they seek to stabilise activity. However, the empirical part of this paper is not an evaluation of monetary policy: it deals with a real lending rate and an official exchange rate, in reduced-form bivariate VARs.

The results do not reveal any robust dynamic relationship common to the three countries. In the Brazilian case, two horizons exclude zero in the baseline VAR(2), but these results are sensitive to the number of lags and to the inclusion of the Covid period. For India and Nigeria, the responses are generally imprecise; the Nigerian results call for additional caution because of the properties of growth and the administered nature of the official exchange rate over part of the period.

These estimates therefore describe dynamic correlations conditional on a specification, not causal effects or central bank reaction functions. The multiplicity of horizons, the absence of a global block and the small sample size reinforce this conclusion. The main contribution of the exercise is to show the limits of an overly direct reading of annual VARs when the aim is to discuss monetary countercyclicality.

An extension intended for journal publication should start from quarterly data, explicitly use policy rates (Selic, repo rate, MPR), include an exogenous global block (US interest rate, dollar, VIX and, for Nigeria, the oil price) and favour either Jordà-style local projections or the direct estimation of reaction functions. These choices would bring the empirical identification closer to the question of countercyclicality itself.

## Appendix 1: Econometric diagnostics and reproducibility

The replication pipeline is documented for Python 3.13.5, pandas 2.2.3, NumPy 2.3.5 and statsmodels 0.14.6. The script checks years, duplicates, missing values and losses at merges; it compares the VAR(1) and VAR(2) over a common window and exports the bootstrap diagnostics. The checks on the Brazil growth-interest rate model (VAR(1) imposed over 2000-2022 and VAR(2) ending in 2019), as well as the reverse responses of the real lending rate to a growth innovation, are now recomputed with the same filtered parametric bootstrap procedure as the main results. The corresponding outputs, including sample sizes and pointwise intervals, are kept in the replication folder.

Tables A1 and A2 make the main diagnostics directly verifiable. For ADF, H0 corresponds to the presence of a unit root; for KPSS, H0 corresponds to stationarity around a constant. The maximum number of ADF lags is set at two and AIC selects the effective lag. The Zivot-Andrews test on Nigerian growth remains a sensitivity check and does not remove the reservations linked to the small sample.

**Table A1 - Stationarity tests on the transformed variables**

| Country | Variable | ADF stat. | ADF p | ADF lags | ADF N | KPSS stat. | KPSS p |
|---|---|---|---|---|---|---|---|
| Brazil | g | -3.352 | 0.013 | 0 | 21 | 0.363 | 0.093 |
| Brazil | Δi | -3.536 | 0.007 | 1 | 20 | 0.073 | ≥ 0.100 |
| Brazil | Δe | -3.151 | 0.023 | 0 | 21 | 0.169 | ≥ 0.100 |
| India | g | -4.334 | < 0.001 | 0 | 21 | 0.119 | ≥ 0.100 |
| India | Δi | -5.685 | < 0.001 | 0 | 21 | 0.107 | ≥ 0.100 |
| India | Δe | -4.101 | < 0.001 | 0 | 21 | 0.252 | ≥ 0.100 |
| Nigeria | g | -2.379 | 0.148 | 0 | 21 | 0.656 | 0.018 |
| Nigeria | Δi | -4.318 | < 0.001 | 2 | 19 | 0.398 | 0.078 |
| Nigeria | Δe | -3.259 | 0.017 | 0 | 21 | 0.222 | ≥ 0.100 |

*Note: g = 100 × Δlog(real GDP); Δi = change in the real interest rate in percentage points; Δe = 100 × Δlog(exchange rate). For Nigerian growth, ADF and KPSS do not support stationarity around a constant; the Zivot-Andrews test suggests a possible break around 2015 (p = 0.005).*

**Table A2 - Model selection and residual diagnostics**

| Country | Variables | Lags | Effective N | BIC | Stable | LB equation | LB lag | Min. p |
|---|---|---|---|---|---|---|---|---|
| Brazil | g, Δi | 2 | 20 | 6.181 | Yes | g | 1 | 0.451 |
| Brazil | g, Δe | 1 | 21 | 7.380 | Yes | g | 3 | 0.580 |
| India | g, Δi | 1 | 21 | 4.667 | Yes | Δi | 3 | 0.386 |

| Country | Variables | Lags | Effective N | BIC | Stable | LB equation | LB lag | Min. p |
|---|---|---|---|---|---|---|---|---|
| India | g, Δe | 1 | 21 | 6.431 | Yes | Δe | 3 | 0.422 |
| Nigeria | g, Δi | 1 | 21 | 6.677 | Yes | Δi | 3 | 0.009 |
| Nigeria | g, Δe | 1 | 21 | 7.521 | Yes | Δe | 3 | 0.379 |

Note: the Ljung-Box tests are computed separately for each equation at lags 1, 2 and 3; the table reports the equation and lag at which the minimum p-value is observed. The warning for the Nigeria growth-real interest rate model comes from the Δi equation at lag 3 ($p = 0.009$). In addition, the multivariate residual whiteness test with a small-sample correction does not reject the hypothesis of non-autocorrelated residuals for any of the six models at the 5% level (p-values from 0.152 to 0.739 depending on the model). These diagnostics remain indicative given the very small sample sizes.

## Appendix 2: Real GDP in constant local currency (2000-2022)

| Year | Brazil | India | Nigeria |
|---|---|---|---|
| 2000 | 2,706,891,619,700 | 69,907,008,757,200 | 72,343,455,024,200 |
| 2001 | 2,744,514,608,900 | 73,279,299,275,800 | 76,624,512,556,900 |
| 2002 | 2,828,317,315,700 | 76,066,825,735,800 | 88,370,403,422,500 |
| 2003 | 2,860,583,579,800 | 82,045,968,414,900 | 94,863,149,255,900 |
| 2004 | 3,025,352,182,400 | 88,546,418,486,000 | 103,638,520,115,900 |
| 2005 | 3,122,227,954,600 | 95,562,332,522,400 | 110,311,303,360,200 |
| 2006 | 3,245,930,273,700 | 103,265,356,587,600 | 116,995,537,398,500 |
| 2007 | 3,442,954,041,300 | 111,176,324,582,100 | 124,706,865,784,600 |
| 2008 | 3,618,344,849,300 | 114,608,002,035,600 | 133,142,627,772,700 |
| 2009 | 3,613,792,537,200 | 123,618,355,749,400 | 143,843,201,044,300 |
| 2010 | 3,885,847,000,000 | 134,122,930,236,000 | 155,358,792,778,800 |
| 2011 | 4,040,287,000,000 | 141,152,737,106,700 | 163,605,119,742,100 |
| 2012 | 4,117,908,023,600 | 148,854,577,483,000 | 170,525,716,392,500 |
| 2013 | 4,241,643,857,400 | 158,360,589,183,700 | 181,902,058,864,500 |
| 2014 | 4,263,019,865,100 | 170,095,469,279,100 | 193,379,567,006,900 |
| 2015 | 4,111,863,267,300 | 183,696,734,680,600 | 198,509,333,815,600 |
| 2016 | 3,977,162,043,300 | 198,863,298,292,600 | 195,299,698,036,200 |
| 2017 | 4,029,774,689,200 | 212,376,821,891,300 | 196,873,592,169,800 |
| 2018 | 4,101,652,440,900 | 226,083,306,267,300 | 200,658,993,617,700 |
| 2019 | 4,151,724,504,300 | 234,835,978,902,900 | 205,090,405,582,400 |
| 2020 | 4,015,682,506,400 | 221,267,802,529,200 | 192,028,406,013,100 |
| 2021 | 4,206,933,577,300 | 242,707,750,910,100 | 194,158,485,984,800 |
| 2022 | 4,333,843,905,000 | 261,176,269,506,000 | 202,543,859,937,900 |

## Appendix 3: Real interest rate used (FR.INR.RINR, 2000-2022)

| Year | Brazil | India | Nigeria |
|---|---|---|---|
| 2000 | 48.5047 | 8.3426 | -1.1409 |
| 2001 | 45.6378 | 8.5914 | 12.1387 |
| 2002 | 48.3404 | 7.9072 | 3.0235 |
| 2003 | 46.4474 | 7.3079 | 9.9357 |
| 2004 | 43.7792 | 4.9101 | -2.6048 |
| 2005 | 44.6352 | 4.8551 | -1.5937 |
| 2006 | 41.2403 | 2.5706 | -5.6280 |
| 2007 | 35.0225 | 5.6818 | 9.1872 |
| 2008 | 35.3668 | 3.7718 | 6.6849 |
| 2009 | 34.7920 | 4.8086 | 18.1800 |
| 2010 | 29.1158 | -1.9839 | 1.0677 |
| 2011 | 32.8335 | 1.3180 | 5.6856 |
| 2012 | 26.5821 | 2.4735 | 6.2248 |
| 2013 | 18.4988 | 3.8660 | 11.2016 |
| 2014 | 22.4037 | 6.6952 | 11.3562 |
| 2015 | 33.8323 | 7.5565 | 13.5962 |
| 2016 | 40.6984 | 6.2327 | 6.6862 |
| 2017 | 41.7138 | 5.3276 | 5.7906 |
| 2018 | 33.1023 | 5.3617 | 6.0560 |
| 2019 | 31.9031 | 6.8949 | 4.5222 |
| 2020 | 21.1972 | 4.1360 | 5.3713 |
| 2021 | 15.0109 | 0.3169 | 1.2277 |
| 2022 | 28.3961 | 1.7046 | 0.9192 |

## Appendix 4: Official exchange rate, local currency per dollar (PA.NUS.FCRF, 2000-2022)

| Year | Brazil | India | Nigeria |
|---|---|---|---|
| 2000 | 1.8294 | 44.9416 | 101.6973 |
| 2001 | 2.3496 | 47.1864 | 111.2313 |
| 2002 | 2.9204 | 48.6103 | 120.5782 |
| 2003 | 3.0775 | 46.5833 | 129.2224 |
| 2004 | 2.9251 | 45.3165 | 132.8880 |
| 2005 | 2.4344 | 44.1000 | 131.2743 |
| 2006 | 2.1753 | 45.3070 | 128.6517 |
| 2007 | 1.9471 | 41.3485 | 125.8081 |
| 2008 | 1.8338 | 43.5052 | 118.5667 |
| 2009 | 1.9994 | 48.4053 | 148.8800 |
| 2010 | 1.7592 | 45.7258 | 150.2975 |
| 2011 | 1.6728 | 46.6705 | 153.8625 |
| 2012 | 1.9531 | 53.4372 | 157.5000 |
| 2013 | 2.1561 | 58.5978 | 157.3117 |
| 2014 | 2.3529 | 61.0295 | 158.5526 |
| 2015 | 3.3269 | 64.1519 | 192.4403 |
| 2016 | 3.4913 | 67.1953 | 253.4920 |
| 2017 | 3.1914 | 65.1216 | 305.7901 |
| 2018 | 3.6538 | 68.3895 | 306.0837 |
| 2019 | 3.9445 | 70.4203 | 306.9210 |
| 2020 | 5.1552 | 74.0996 | 358.8108 |
| 2021 | 5.3944 | 73.9180 | 401.1520 |
| 2022 | 5.1640 | 78.6045 | 425.9792 |